\documentclass[proceedings]{JHEP} 

\usepackage{epsfig}			

\newbox\mybox
\newcommand\fverb{\setbox\mybox=\hbox\bgroup\verb}
\newcommand\fverbdo{\egroup\medskip\noindent\fbox{\unhbox\mybox}\ }
\newcommand\fverbit{\egroup\item[\fbox{\unhbox\mybox}]}

\font\beeg=cmr17 scaled 1600		
\newcommand\init[1]{\setbox\mybox=\hbox{{\beeg #1}~}%
		   \noindent\global\hangindent=\wd\mybox\global\hangafter-2%
		   \sc\smash{\llap {\lower 13.2pt \box\mybox}}}

\title{Relativistic quark-antiquark potential and heavy quarkonium mass spectra}

\author{D.\ Ebert$^\dag$, R.\ N.\ Faustov$^\ddag$, V.\ O.\ Galkin$^\ddag$\\
$^\dag$ Institut f\"ur Physik, Humboldt--Universit\"at zu Berlin,
Invalidenstr.110, D-10115 Berlin, Germany\\
$^\ddag$ Russian Academy of Sciences,
Scientific Council for Cybernetics,
Vavilov Street 40, Moscow 117333, Russia}

\conference{Heavy Quark Physics 5, Dubna, Russia, 6-8 April 2000}

\abstract{A general approach to accounting for retardation effects in the 
long-range (confining) part of the quark-antiquark potential is presented.
The charmonium and bottomonium mass spectra are calculated with the
systematic account of  relativistic and retardation effects and
the one-loop radiative corrections. A good fit to available experimental
data on the mass spectra is obtained.}

\begin{document} 

The relativistic properties of the quark-anti\-quark interaction potential
play an important role in analysing different static and dynamical 
characteristics of heavy me\-sons. The Lorentz-structu\-re of the confining 
quark-anti\-quark interaction is of particular interest. In the literature 
there is no consent on this item. For a 
long time the scalar confining kernel has been considered to be the
most appropriate one \cite{scal}. The main argument in favour of this
choice is based on the nature of the heavy quark spin-orbit potential.
The scalar potential gives a vanishing long-range magnetic 
interaction, which is in agreement with the flux tube picture
of quark confinement of~\cite{buch}, and allows to get the fine
structure for heavy quarkonia in accord with experimental data. 
However, the calculations of electroweak decay rates of heavy mesons 
with a scalar confining potential alone yield results which are in  worse 
agreement with data than with a vector potential \cite{gf}. 
The  radiative
$M1$-transitions in quarkonia such as e.~g. $J/\psi\to \eta_c
\gamma$ are the most sensitive
to the Lorentz-structure of the confining potential. 
The relativistic corrections for 
these decays arising from vector and scalar potentials have different
signs \cite{gf}. In particular, as it has been 
shown in ref.~\cite{gf}, agreement
with experiments for these decays can be achieved only for a specific mixture
of vector and scalar potentials. In this context, it is worth noting,
that the recent study of the $q\bar q$ 
interaction in the Wilson loop approach \cite{bv1} indicates that
it cannot be considered as purely  scalar. Moreover, the found
structure of spin-independent relativistic corrections is not 
compatible with a scalar potential. A similar conclusion
has been obtained in ref.~\cite{ss} on the basis of a Foldy-Wouthuysen 
reduction of the full Coulomb gauge Hamiltonian of QCD. There, the 
Lorentz-structure of the confinement has been found to be of vector
nature. The scalar nature of spin splittings in heavy quarkonia
in this approach is dynamically generated through the interaction
with collective gluonic degrees of freedom. Thus we see that while the
spin-dependent structure of $q\bar q$ interaction is well established
now,  the spin-in\-de\-pen\-dent part is still controversial in the 
literature. 
The uncertainty in the Lorentz-structure of the confining interaction 
complicates the account of retardation corrections since the relativistic
reconstruction of the static confining potential is not unique.
Here we present the generalized prescription of such reconstruction and 
discuss its implications for the heavy quarkonium mass spectra. 

In our preceding papers we have developed the relativistic quark model based
on the quasipotential approach. A meson is described by the wave
function of the bound quark-antiquark state, which satisfies the
quasipotential equation \cite{3} of the Schr\"odinger type \cite{4}\newpage
\begin{eqnarray}
\label{quas}
&&{\left(\frac{b^2(M)}{2\mu_{R}}-\frac{{\bf
p}^2}{2\mu_{R}}\right)\Psi_{M}({\bf p})}\cr
&& \quad =\int\frac{d^3 q}{(2\pi)^3}
 V({\bf p,q};M)\Psi_{M}({\bf q}),\qquad
\end{eqnarray}
where the relativistic reduced mass is
\begin{equation}
\mu_{R}=\frac{E_aE_b}{E_a+E_b}=\frac{M^4-(m^2_a-m^2_b)^2}{4M^3},
\end{equation}
and $E_a$, $E_b$ are given by
\begin{equation}
\label{ee}
E_a=\frac{M^2-m_b^2+m_a^2}{2M}, \quad E_b=\frac{M^2-m_a^2+m_b^2}{2M}.
\end{equation}
Here $M=E_a+E_b$ is the meson mass, $m_{a,b}$ are the masses of light
and heavy quarks, and ${\bf p}$ is their relative momentum.  
In the centre of mass system the relative momentum squared on mass shell 
reads
\begin{equation}
{b^2(M) }
=\frac{[M^2-(m_a+m_b)^2][M^2-(m_a-m_b)^2]}{4M^2}.
\end{equation}

The kernel 
$V({\bf p,q};M)$ in Eq.~(\ref{quas}) is the quasi\-po\-ten\-tial operator of
the quark-antiquark interaction. It is constructed with the help of the
off-mass-shell scattering amplitude, projected onto the positive
energy states. 
Constructing the quasi\-po\-ten\-tial of the quark-antiquark interaction 
we have assumed that the effective
interaction is the sum of the usual one-gluon exchange term with the mixture
of long-range vector and scalar linear confining potentials, where
the vector confining potential
contains the Pauli interaction. The quasi\-po\-ten\-tial is then defined by
\cite{mass}  
\begin{eqnarray}
\label{qpot}
&&V({\bf p,q};M)=\bar{u}_a(p)
\bar{u}_b(-p)\Bigg\{\frac{4}{3}\alpha_sD_{ \mu\nu}({\bf
k})\gamma_a^{\mu}\gamma_b^{\nu}\cr
& &\quad +V_V({\bf k})\Gamma_a^{\mu}
\Gamma_{b;\mu}+V_S({\bf
k})\Bigg\}u_a(q)u_b(-q),\qquad
\end{eqnarray}
where $\alpha_S$ is the QCD coupling constant, $D_{\mu\nu}$ is the
gluon propagator in the Coulomb gauge
and ${\bf k=p-q}$; $\gamma_{\mu}$ and $u(p)$ are 
the Dirac matrices and spinors
with $\epsilon(p)=\sqrt{p^2+m^2}$.
The effective long-range vector vertex is
given by
\begin{equation}
\label{kappa}
\Gamma_{\mu}({\bf k})=\gamma_{\mu}+
\frac{i\kappa}{2m}\sigma_{\mu\nu}k^{\nu},
\end{equation}
where $\kappa$ is the Pauli interaction constant characterizing the
anomalous chromomagnetic moment of quarks. Vector and
scalar confining potentials in the nonrelativistic limit reduce to
\begin{equation}
\label{vlin}
V_V(r)=(1-\varepsilon)Ar+B,\quad
V_S(r) =\varepsilon Ar,
\end{equation}
reproducing 
\begin{equation}
\label{nr}
V_{\rm conf}(r)=V_S(r)+V_V(r)=Ar+B,
\end{equation}
where $\varepsilon$ is the mixing coefficient. 

The retardation contribution to the one-gluon exchange part of the $q\bar q$ 
potential is well known.
For the confining part of the $q\bar q$ potential the retardation contribution is much
more indefinite. It is a 
consequence of our poor knowledge of the confining potential
especially in what concerns its relativistic properties: the Lorentz structure
(scalar, vector, etc.) and the dependence on the covariant variables such as 
$k^2=k_0^2-{\bf k}^2$. Nevertheless we can perform some general considerations and then apply them to a particular case of the linearly rising potential. To this end we
note that for any nonrelativistic potential $V(-{\bf k}^2)$ the simplest relativistic
generalization is to replace it by $V(k_0^2-{\bf k}^2)$.

In the case of the Lorentz-vector confining potential we can use the same approach
as for the one-gluon exchange
even with more general vertices containing the Pauli terms, since the
mass-shell vector currents are conserved here as well. It is possible to introduce
alongside with the ``diagonal gauge" the so-called ``instantaneous gauge'' 
\cite{ch} which 
is the generalization of the Coulomb gauge:
\begin{eqnarray}
\label{vpr}
&&V_V(k_0^2-{\bf k}^2)\bar u_{a}({\bf p})\bar u_b(-{\bf p})
\Gamma^\mu_a\Gamma_{b\mu}
u_a({\bf q})u_b({-\bf q})\cr
&&\quad=\bar u_{a}({\bf p})\bar u_b(-{\bf p})\biggl\{V_V(-{\bf k}^2){\Gamma_a^0
\Gamma_b^0}\cr
&&\quad
 -\bigl[V_V(-{\bf k}^2){\mathbf{\Gamma}_a\cdot \mathbf{\Gamma}_b}
+V'_V(-{\bf k}^2)(\mathbf{\Gamma}_a \cdot
{\bf k})\cr
&&\quad\times(\mathbf{\Gamma}_b\cdot{\bf k})\bigr]\biggr\}
 u_a({\bf q})u_b({-\bf q}),
\end{eqnarray}
where
$$V_V(k_0^2-{\bf k}^2)\cong V_V(-{\bf k}^2)+k_0^2V'_V(-{\bf k}^2)$$
and 
\begin{equation}
\label{k0s}
k_0^2=(\epsilon_a({\bf p})-\epsilon_a({\bf q}))(\epsilon_b({\bf q})-\epsilon_b({\bf p}))
\cong-\frac{({\bf p}^2-{\bf q}^2)^2}{4m_am_b}
\end{equation}
with the correct Dirac limit in which the retardation contribution vanishes when
one of the particles becomes infinitely heavy  \cite{om}.

For the case of the Lorentz-scalar potential we can make the same expansion 
in $k_0^2$, which yields
\begin{equation}
\label{vse}
V_S(k_0^2-{\bf k}^2)\cong V_S(-{\bf k}^2)+k_0^2V'_S(-{\bf k}^2).
\end{equation}
But in this case we have no reasons to fix $k_0^2$ in the only way (\ref{k0s}).
The other possibility is to take a half sum instead of a symmetrized product,
namely to set (see e.~g. \cite{gromes,om})
\begin{eqnarray}
\label{k0hs}
k_0^2&=&\frac12\left[(\epsilon_a({\bf p})-\epsilon_a({\bf q}))^2+
(\epsilon_b({\bf q})-\epsilon_b({\bf p}))^2\right]\cr
&\cong&
\frac18({\bf p}^2-{\bf q})^2\left(\frac{1}{m_a^2}+\frac{1}{m_b^2}\right).
\end{eqnarray}
The Dirac limit is not fulfilled by this choice, but this 
cannot serve as a decisive
argument. Thus the most general expression for the energy transfer squared,
which incorporates both possibilities (\ref{k0s}) and (\ref{k0hs}) has the form
\begin{eqnarray}
\label{k00}
&&k_0^2=\lambda(\epsilon_a({\bf p})-\epsilon_a({\bf q}))
(\epsilon_b({\bf q})-\epsilon_b({\bf p}))
+(1-\lambda)\cr
&&\qquad\times\frac12\left[(\epsilon_a({\bf p})-\epsilon_a({\bf q}))^2+
(\epsilon_b({\bf q})-\epsilon_b({\bf p}))^2\right]\cr
&&\quad\cong-\lambda\frac{({\bf p}^2-{\bf q}^2)^2}{4m_am_b}\cr
&&\qquad+(1-\lambda)\frac18({\bf p}^2-{\bf q})^2\left(\frac{1}{m_a^2}+\frac{1}{m_b^2}\right),
\end{eqnarray}
where $\lambda$ is the mixing parameter.

Thus the spin-independent part of $q\bar q$ potential with the account of 
retardation corrections takes the form:
\begin{eqnarray}
\label{spind}
&&\!\!\!\! V_{\rm SI}(r)=V_C(r)+V_{\rm conf}(r) + V_{\rm VD}(r) \phantom{qqqqqqqqqqqq}\cr
&&\!\!\!\! +\frac18\left(\frac{1}{m_a^2}+
\frac{1}{m_b^2}\right)\Delta\big[V_C(r) 
 +(1+2\kappa)V_V(r)\big],
\end{eqnarray}
where the velocity-dependent part
\begin{eqnarray}
\label{vd}
&&V_{\rm VD}(r)= V_{\rm VD}^C(r)+V_{\rm VD}^V(r)+V_{\rm VD}^S(r),\\
&&V_{\rm VD}^C(r)=\frac{1}{2m_am_b}\left\{V_C(r)\left[{\bf p}^2
+\frac{({\bf p\cdot r})^2}{r^2}\right]\right\}_W  \cr
&&V_{\rm VD}^V(r)=\frac{1}{m_am_b}\left\{V_V(r){\bf p}^2\right\}_W\cr
&&+\frac14\Biggl[
(1-\lambda_V)\left(\frac{1}{m_a^2}+\frac{1}{m_b^2}\right)-\frac{2\lambda_V}{m_am_b}
\Biggr]\cr
&&\times \left\{V_V(r){\bf p}^2+V_V'(r)\frac{({\bf p\cdot r})^2}{r}\right\}_W,\cr
&&V_{\rm VD}^S(r)=\frac{1}{2}\left(\frac{1}{m_a^2}
+\frac{1}{m_b^2}\right)\left\{V_V(r){\bf p}^2\right\}_W\cr
&&+
\frac14\Biggl[
(1-\lambda_S)\left(\frac{1}{m_a^2}+\frac{1}{m_b^2}\right)-\frac{2\lambda_S}{m_am_b}
\Biggr]\cr
&&\times \left\{V_V(r){\bf p}^2+V_V'(r)\frac{({\bf p\cdot r})^2}{r}\right\}_W
 \nonumber 
\end{eqnarray}
and $\{\dots\}_W$ denotes the Weyl ordering of operators.
Making the natural decomposition
\begin{eqnarray}
\label{vdrel}
&&\!\!\!\!\!V_{\rm VD}(r)=\frac{1}{m_am_b}\left\{{\bf p}^2V_{bc}(r)+\frac{({\bf p
\cdot r})^2}{r^2}V_c(r)\right\}_W \cr
&&\!\!\!\!\!+\left(\frac{1}{m_a^2}+\frac{1}{m_b^2}\right)\left\{{\bf p}^2 V_{de}
(r) -\frac{({\bf p\cdot r})^2}{r^2}V_e(r)\right\}_W
\end{eqnarray}
we obtain \cite{mass1} for the corresponding structures with $\lambda_V=1$ and
including one-loop radiative corrections in $\overline {MS}$
renormalization scheme:
\begin{eqnarray}
\label{potcoef}
&&V_C(r)=-\frac43\frac{\bar \alpha_V(\mu^2)}{r}  -\frac43\frac{\beta_0
\alpha_s^2(\mu^2)}{2\pi}\frac{\ln(\mu r)}{r}, \cr
&&V_{bc}(r)=-\frac23\frac{\bar \alpha_V(\mu^2)}{r}  -\frac23\frac{\beta_0
\alpha_s^2(\mu^2)}{2\pi}\frac{\ln(\mu r)}{r}\cr
&& \quad+\left(\frac{1-\varepsilon}{2}
-\frac{\varepsilon\lambda_S}{2}\right)Ar+B,\cr
&&V_c(r)=-\frac23\frac{\bar \alpha_V(\mu^2)}{r}  -\frac23\frac{\beta_0
\alpha_s^2(\mu^2)}{2\pi}\cr
&&\quad\times
\biggl[\frac{\ln(\mu r)}{r}
-\frac1r\biggr]
-\left(\frac{1-\varepsilon}{2}+\frac{\varepsilon\lambda_S}{2}\right)Ar,\cr
&&V_{de}(r)= -\frac{\varepsilon}{4}(1+\lambda_S)Ar+B,\cr
&&V_e(r)= -\frac{\varepsilon}{4}(1-\lambda_S)Ar,
\end{eqnarray}
where
\begin{eqnarray*}
\bar\alpha_V(\mu^2)&=&\alpha_s(\mu^2)\left[1+\left(\frac{a_1}{4}
+\frac{\gamma_E\beta_0}{2}\right)\frac{\alpha_s(\mu^2)}{\pi}\right],\\
a_1&=&\frac{31}{3}-\frac{10}{9}n_f,\quad
\beta_0=11-\frac23n_f.\nonumber
\end{eqnarray*}
Here $n_f$ is a number of flavours and $\mu$ is a renormalization scale.
It is easy to check that the exact  Barchielli, Brambilla, Prosperi 
relations \cite{bbp} following from the Lorentz
invariance of the Wilson loop 
\begin{eqnarray}
\label{re}
&&V_{de}-\frac12 V_{bc}+\frac14(V_C+V_0)=0, \cr
&&V_e+\frac12 V_c+\frac{r}{4}\frac{{\rm d}(V_C+ V_0)}{{\rm d} r}=0
\end{eqnarray}
are exactly satisfied.

The expression for spin-dependent part of the quark-antiquark 
potential  with the inclusion
of radiative corrections  can be found in ref.~\cite{mass1}. Now we can calculate 
the mass spectra of heavy quarkonia with the account
of all relativistic corrections (including retardation effects) of order $v^2/c^2$
and one-loop radiative corrections. For this purpose we substitute the quasipotential
which is a sum of the spin-independent   and spin-dependent 
parts into the quasipotential equation. Then we multiply the resulting
expression from the left by the quasipotential wave function of a bound state and
integrate with respect to the relative momentum. Taking into account the accuracy
of the calculations, we can use for the resulting matrix elements the wave functions of
Eq.~(\ref{quas}) with the static potential 
\begin{equation}
V_{\rm NR}(r)=-\frac43\frac{\bar \alpha_V(\mu^2)}{r} +Ar+B.
\end{equation}
As a result we obtain the mass formula ($m_a=m_b=m$)
\begin{eqnarray}
\label{mform}
&&\frac{b^2(M)}{2\mu_R}=W+\langle a\rangle\langle{\bf L}\cdot{\bf S}\rangle +\langle b\rangle \langle\biggl[-({\bf S}_a\cdot {\bf S}_b)\cr
&&+ \frac{3}{r^2}
({\bf S}_a\cdot {\bf r})
({\bf S}_b\cdot {\bf r})\biggr] \rangle
+\langle c\rangle \langle{\bf S}_a\cdot {\bf S}_b\rangle,
\end{eqnarray}
where the first term on the right-hand side of the mass formula contains
all spin-independent contributions, the second term describes the spin-orbit
interaction, the third term is responsible for the tensor interaction, while the
last term gives the spin-spin interaction.

To proceed further we need to discuss the parameters of our model. There is the
following set of parameters: the quark masses ($m_b$ and $m_c$), the QCD
constant $\Lambda$ and renormalization point $\mu$ 
in the short-range part of the $Q\bar Q$ potential,
the slope $A$ and intercept $B$ of the
linear confining potential (\ref{nr}), the mixing
coefficient $\varepsilon$ (\ref{vlin}), the long-range anomalous chromomagnetic
moment $\kappa$ of the quark (\ref{kappa}), and the mixing parameter $\lambda_S$
in the retardation correction for the scalar confining potential.   
We can fix the values of the parameters $\varepsilon=-1$ and
$\kappa=-1$ from the consideration of radiative decays \cite{gf} and comparison
of the heavy quark expansion in our model \cite{fg,exc}
with the predictions of the heavy quark effective theory. We  fix the slope of
the linear confining potential $A=0.18$~GeV$^2$ which is a rather adopted
value. In order to reduce the number of independent parameters we assume
that the renormalization scale $\mu$ in the strong 
coupling constant $\alpha_s(\mu^2)$
is equal to the quark mass. We also varied the quark masses
in a reasonable range for the constituent quark masses. The numerical analysis
and comparison with experimental data
lead to the following values of our model parameters:
$m_c=1.55~{\rm GeV}, \, m_b=4.88~{\rm GeV},\, 
A=0.18~{\rm GeV}^2, \, B=-0.16~{\rm GeV},\, \mu=m_Q,\, 
\Lambda=0.178~{\rm GeV}, \,\varepsilon=-1, \, \kappa=-1, \, \lambda_S=0. $
The quark masses $m_{c,b}$ have usual values for constituent quark models
and coincide with those chosen in our previous analysis \cite{mass}. The above
value of the retardation parameter $\lambda_S$ for the
 scalar confining potential
coincides with the minimal area low and flux tube models \cite{bv}, with
lattice results \cite{bali} and Gromes suggestion \cite{gromes}. The found
value for the QCD parameter $\Lambda$ gives the following
values for the strong coupling constants $\alpha_s(m_c^2)\approx 0.32$ and
$\alpha_s(m_b^2)\approx 0.22$.

\TABLE[t]{
\caption{Charmonium mass spectrum. }
\label{charm}
\begin{tabular}{cccc}
\hline
State & Particle &  Theory & Experiment \cite{pdg}\\
\hline
$1^1S_0$& $\eta_c$ & 2.979 & 2.9798 \\
$1^3S_1$& $J/\Psi$ & 3.096 & 3.09688 \\
$1^3P_0$& $\chi_{c0}$ & 3.424 & 3.4173 \\
$1^3P_1$& $\chi_{c1}$ & 3.510 & 3.51053 \\
$1^3P_2$& $\chi_{c2}$ & 3.556 & 3.55617 \\
$2^1S_0$& $\eta_c'$ & 3.583 & 3.594 \\
$2^3S_1$& $\Psi'$     & 3.686 & 3.686 \\ 
$1^3D_1$&  & 3.798 & $3.7699$ \\
$1^3D_2$&  & 3.813 & \\
$1^3D_3$&  & 3.815 & \\
$2^3P_0$& $\chi'_{c0}$ & 3.854 & \\
$2^3P_1$& $\chi'_{c1}$ & 3.929 & \\
$2^3P_2$& $\chi'_{c2}$ & 3.972 & \\
$3^1S_0$& $\eta_c''$ & 3.991 & \\
$3^3S_1$& $\Psi''$     & 4.088 & $4.040$ \\
$2^3D_1$&  & 4.194 & $4.159$ \\
$2^3D_2$&  & 4.215 & \\
$2^3D_3$&  & 4.223 & \\ 
\hline
\end{tabular}

}
\TABLE[t]{
\caption{Bottomonium mass spectrum. }
\label{bottom}
\begin{tabular}{cccc}
\hline
State & Particle &  Theory & Experiment \cite{pdg} \\
\hline
$1^1S_0$& $\eta_b$ & 9.400 &  \\
$1^3S_1$& $\Upsilon$ & 9.460 & 9.46037 \\
$1^3P_0$& $\chi_{b0}$ & 9.864 & 9.8598 \\
$1^3P_1$& $\chi_{b1}$ & 9.892 & 9.8919 \\
$1^3P_2$& $\chi_{b2}$ & 9.912 & 9.9132 \\
$2^1S_0$& $\eta_b'$ & 9.990 & \\
$2^3S_1$& $\Upsilon'$  & 10.020 & 10.023 \\ 
$1^3D_1$&  & 10.151 &   \\
$1^3D_2$&  & 10.157 &  \\
$1^3D_3$&  & 10.160 & \\
$2^3P_0$& $\chi'_{b0}$ & 10.232 & 10.232  \\
$2^3P_1$& $\chi'_{b1}$ & 10.253 & 10.2552  \\
$2^3P_2$& $\chi'_{b2}$ & 10.267 & 10.2685  \\
$3^1S_0$& $\eta_b''$ & 10.328 & \\
$3^3S_1$& $\Upsilon''$ & 10.355 & 10.3553 \\
$2^3D_1$&  & 10.441 &   \\
$2^3D_2$&  & 10.446 &  \\
$2^3D_3$&  & 10.450 &  \\ 
$3^3P_0$& $\chi''_{b0}$ & 10.498 &   \\
$3^3P_1$& $\chi''_{b1}$ & 10.516 &   \\
$3^3P_2$& $\chi''_{b2}$ & 10.529 &   \\
$4^1S_0$& $\eta_b'''$ & 10.578 &  \\
$4^3S_1$& $\Upsilon'''$ & 10.604 & 10.580 \\
\hline
\end{tabular}
}

The results of our numerical calculations of the mass spectra of charmonium and
bottomonium (in GeV) are presented 
in Tables~\ref{charm} and \ref{bottom}. We see that
the calculated masses agree with experimental 
values within few MeV and this difference
is compatible with the estimates of the higher order corrections in $v^2/c^2$ and
$\alpha_s$. The model reproduces correctly both the positions of the centres of
gravity of the levels and their fine and hyperfine splitting.
Note that the good    
agreement of the calculated mass spectra with experimental data is achieved 
by systematic accounting for all relativistic corrections 
(including retardation corrections) of 
order $v^2/c^2$, both spin-dependent and 
spin-independent ones, while in most 
of potential models only the spin-depen\-dent
corrections are included.

The calculated mass spectra of charmonium and
bottomonium are  close to the results of
our previous calculation \cite{mass} where 
retardation effects in the confining potential and radiative corrections 
to the one-gluon exchange potential were not taken into account. 
Both calculations
give close values for the experimentally measured sta\-tes as well as for
the yet unobserved ones. The inclusion of radiative corrections allowed to get
better results for the fine splittings of quarkonium states. 
Thus we can conclude
from this comparison that the
inclusion of retardation effects and spin-independent
one-loop radiative corrections resulted only in the slight 
shift ($\approx 10\%$) in the value of 
the QCD parameter $\Lambda$ and an approximately two-fold decrease of the 
constant $B$.
Such changes of parameters almost do not influence the wave functions. As
a result the decay matrix elements involving heavy quarkonium states remain
mostly unchanged.

We are grateful to the organizers for the nice meeting and stimulating discussions.
Two of us (R.N.F and V.O.G.) were supported in part  by {\it Russian Foundation for
Fundamental Research} under Grant No.\ 00-02-17768. 


\begin{thebibliography}{999}
\bibitem{scal} H.J. Schnitzer, \prl{35}{1975}{1540}; 
W. Lucha, F.F. Sch\"oberl and D. Gromes, \prep{200}{1991}{127};  
Yu.A. Simonov, {\it Phys. Usp.} {\bf 39} (1996) 313.
\bibitem{buch} W. Buchm\"uller, \plb{112}{1982}{479}.
\bibitem{gf} V.O. Galkin and R.N. Faustov, 
 \sjnp{44}{1986}{1023}; V.O. Galkin, A.Yu. Mishurov and R.N. Faustov,  
\sjnp{51}{1990}{705}.
\bibitem{bv1} N. Brambilla and A. Vairo, \plb{407}{1997}{167}.
\bibitem{ss} A.P. Szczepaniak and E.S. Swanson, \prd{55}{1997}{3987}.
\bibitem{3} A.A. Logunov and A.N. Tavkhelidze, \nc{29}{1963}{380}.
\bibitem{4} A.P. Martynenko and R.N. Faustov,{\it Theor.
Math. Phys.} {\bf 64} (1985) 765.
\bibitem{mass} V.O. Galkin, A.Yu. Mishurov and R.N. Faustov, 
\sjnp{55}{1992}{1207}.
\bibitem{ch} W. Celmaster and F.S. Henyey, \prd{17}{1978}{3268}.
\bibitem{om} M.G. Olson and K.J. Miller, \prd{28}{1983}{674}.
\bibitem{gromes} D. Gromes, \npb{131}{1977}{80}.
\bibitem{mass1} D. Ebert, R.N. Faustov and V.O. Galkin, 
\hepph{9911283}, 
{\it Phys. Rev. }{\bf D} in press.
\bibitem{bbp} A. Barchielli, N. Brambilla and G.M. Prosperi, \nc{103}{1990}{59}.
\bibitem{fg} R.N. Faustov and V.O. Galkin, \zpc{66}{1995}{119}.
\bibitem{exc} D. Ebert, R.N. Faustov and V.O. Galkin, \plb{454}{1998}{365};
\prd{61}{2000}{014016}; \prd{62}{2000}{014032}.
\bibitem{bv} N. Brambilla and A. Vairo, \prd{55}{1997}{3974}.
\bibitem{bali} G.S. Bali, A. Wachter and K. Schilling, \prd{\bf 56}{1997}{2566}.
\bibitem{pdg} C. Caso {\it et al.}, Particle Data Group, \epjc{3}{1998}{1}.
\end{thebibliography}
\end{document}